\documentclass[aps,pre,twocolumn,superscriptaddress]{revtex4}

\usepackage[utf8x]{inputenc}
\usepackage{ae} 
\usepackage[T1]{fontenc}
\usepackage[english]{babel}
\usepackage{amsmath}
\usepackage{amsfonts}
\usepackage{amssymb}
\usepackage{array}
\usepackage{graphics}
\usepackage{graphicx}
\usepackage{epsfig}
\usepackage{latexsym}
\usepackage{textcomp}
\usepackage{gensymb}
\usepackage{verbatim}
\usepackage{subfigure}
\usepackage[retainorgcmds]{IEEEtrantools}
\usepackage{color}

\newcommand{\rd}{\mathrm{d}}
\newcommand{\nl}{\nonumber \\}
\newcommand{\mc}{\IEEEeqnarraymulticol}
\newcommand{\la}{\langle}
\newcommand{\ra}{\rangle}
\newcommand{\eps}{\varepsilon}

\newcommand{\kpzh}{h}

\begin{document}

\title{Extrema statistics in the dynamics of a non-Gaussian random field}
\date{}

\author{T.H. Beuman}
  \affiliation{Instituut-Lorentz for Theoretical Physics, Leiden University, NL 2333 CA Leiden, The Netherlands}
\author{A.M. Turner}
  \affiliation{Institute for Theoretical Physics, Universiteit van Amsterdam, NL 1090 GL Amsterdam, The Netherlands}
\author{V. Vitelli}
  \email{vitelli@lorentz.leidenuniv.nl}
  \affiliation{Instituut-Lorentz for Theoretical Physics, Leiden University, NL 2333 CA Leiden, The Netherlands}

\date{\today}

\begin{abstract}



When the equations that govern the dynamics of a random field are nonlinear, the field can develop with time non-Gaussian statistics even if its initial condition is Gaussian. Here, we provide a general framework for calculating the effect of the underlying nonlinear dynamics on the relative densities of maxima and minima of the field. Using this simple geometrical probe, we can identify the size of the non-Gaussian contributions in the random field, or alternatively the magnitude of the nonlinear terms in the underlying equations of motion. We demonstrate our approach by applying it to an initially Gaussian field that evolves according to the deterministic KPZ equation, which models surface growth and shock dynamics.

\end{abstract}

\maketitle


Random fields that undergo a time evolution according to a nonlinear dynamical equation often develop non-Gaussian statistics that provide clues about the details of the underlying microscopic mechanisms.
Consider for example a gas-liquid phase transition. In the early stages, there are many randomly small volumes in which all the molecules are in the same phase, distributed randomly. Over time, these volumes will grow and merge, thereby gradually replacing the Gaussian disorder with structure \cite{cite_Bray}.



Even if the initial condition of a random field is Gaussian, the dynamics will typically generate a non-Gaussian component in the field that we wish to quantify and track with time. The standard approach to detect and measure non-Gaussianities is to employ higher-order correlation functions.
In this work, we adopt a geometric approach to measuring the non-Gaussian component of a scalar field $h(\vec{r}, t)$: we interpret it as a height function describing an evolving surface, and study its geometry. Gaussian surfaces have certain general geometric and topological properties \cite{cite_Dennis1, cite_Longuet2, cite_Berry, cite_Longuet1, cite_Dennis2}. For example, the number of maxima exactly balances the number of minima. A random surface that does not exhibit this property is then guaranteed to have non-Gaussian statistics \cite{cite_paper1, cite_paper2}.



In previous articles \cite{cite_paper1, cite_paper2} we studied fields that are local functions of a given Gaussian, i.e.\ of the form $h(\vec{r}) = H(\vec{r}) + f_{NL}(H(\vec{r}))$, where $H$ is a Gaussian field and $f_{NL}$ a nonlinear function. In this scheme, the perturbed height $h$ at any point $\vec{r}$ is a function only of the original height $H(\vec{r})$ at the same point. In this paper, we move to the general case of \emph{nonlocal} perturbations, which e.g.\ include a dependence on $\nabla H$, thereby introducing a mixing between the field values at different points.

Such a nonlocal non-Gaussianity can arise in a broad range of physical contexts, for example as the result of nonlinear diffusion. For concreteness, consider a diffusion equation of the general form
\begin{equation}
  \frac{\partial h(\vec{r}, t)}{\partial t}  =  D \nabla^2 h(\vec{r}, t) + f_{NL}(h, \nabla h),
  \label{eq_diff_eqn}
\end{equation}
where $f_{NL}$ is any nonlinear function. If we let $h$ be a Gaussian field at $t=0$, then non-Gaussianities will emerge as a consequence of the last term; if we would omit this term, we retrieve the heat equation, which would preserve the Gaussianity of $h$ for all $t>0$.
A variety of known diffusion equations has this general form. For instance, when $f_{NL}$ takes the form $-h^2$ we get Fisher's equation, which can be used as a model to describe the growth and saturation of a population. Another example is the Cahn-Hilliard equation for the development of order after a phase transition \cite{cite_Bray}. Several models of structure formation, in both condensed matter \cite{cite_Chaikin} and cosmology \cite{cite_Dodelson}, also belong to this class.

To illustrate our general result, we apply it to the case of a field obeying the deterministic KPZ equation \cite{cite_KPZ}, for which $f_{NL} = \frac{\lambda}{2} (\nabla h)^2$. This equation is often used to model the height profile of a growing surface. A field that starts out as a Gaussian field will acquire non-Gaussian characteristics as time progresses. We use our formula to quantify the resulting effect on the relative difference in densities of maxima and minima. This allows to back up the non-Gaussian component in $h$, or alternatively, to deduce what the nonlinear coefficient $\lambda$ is. We verify the analytical predictions by comparing them with results from computer simulations.

The outline of this paper is as follows. In section~\ref{sec_nongsn_fields} we determine a general expression for the imbalance between maxima and minima for a non-Gaussian field. This is applied to the KPZ equation in section~\ref{sec_kpz}. Finally, section~\ref{sec_conclusions} summarizes our conclusions.

		\section{Non-Gaussian fields}
	\label{sec_nongsn_fields}

A homogeneous and isotropic Gaussian field is defined in terms of its Fourier components as
\begin{equation}
  H(\vec{r})  =  \sum_{\vec{k}} A(k) \cos(\vec{k} \cdot \vec{r} + \phi_{\vec{k}}).
  \label{eq_gaussian}
\end{equation}
The phases $\phi_{\vec{k}}$ are independent random variables, uniformly distributed between $0$ and $2\pi$. The \emph{amplitude spectrum} $A(k)$ depends only on the magnitude of the wave vector $\vec{k}$ and encodes the special features of the Gaussian field under consideration. An alternative approach is to express the amplitude spectrum in terms of its \emph{moments}, according to
\begin{equation}
  K_n  =  \sum_{\vec{k}} \tfrac12 A(k)^2 k^n.
  \label{eq_moments}
\end{equation}
For convenience, we will consider $H$ to be normalized, such that $K_0 = \la H^2 \ra = 1$, see ref.~\cite{cite_paper1} for more details.

In what follows, we concentrate on homogeneous and isotropic fields $h(\vec{r})$, which we assume to be in the form of a Gaussian $H(\vec{r})$ with the addition of a perturbation. Unlike refs.~\cite{cite_paper1, cite_paper2}, we will not restrict ourselves to a perturbation of the local kind, i.e.\ where the perturbation at any point $\vec{r}$ is a function of $H(\vec{r})$ only. We will now also accommodate perturbations which depend on $\vec{\nabla} H$ for instance, or evolve over time. Such perturbations introduce a mixing between the values of the field at different points, which we will designate as \emph{nonlocal} perturbations.

We will investigate the effect of a perturbation on the densities of maxima and minima. A maximum (minimum) $\vec{r_0}$ of $h$ is defined by the condition $h_x(\vec{r_0}) = h_y(\vec{r_0}) = 0$, along with the inequalities $h_{xx}(\vec{r_0}) h_{yy}(\vec{r_0}) - h_{xy}(\vec{r_0})^2 > 0$ (if this were negative, $\vec{r_0}$ would be a saddle point) and $h_{xx}(\vec{r_0})$, $h_{yy}(\vec{r_0})$ negative (positive); note that the first condition implies that $h_{xx}(\vec{r_0})$ and $h_{yy}(\vec{r_0})$ have the same sign. The $x$ and $y$ subscripts indicate derivatives with respect to the coordinates of the two-dimensional plane.

The general procedure that we use is very similar to the one in \cite{cite_paper2} and is as follows: we consider a fixed point $\vec{r_0}$ -- due to the homogeneity of $h$, the analysis will not depend on this choice. We determine the joint probability distribution of $h_x$, $h_y$, $h_{xx}$, $h_{yy}$ and $h_{xy}$, since these stochastic variables are the ingredients from which maxima and minima are defined, as outlined above. This distribution can be determined via the generating function, which in turn can be constructed by determining the relevant cumulants involving the five stochastic variables. Once the probability distribution is obtained, we set $h_x = h_y = 0$ and integrate the second derivatives over the region defining a minimum (maximum) in order to get the density of minima (maxima).

As we did in \cite{cite_paper2}, we transform to another coordinate system, based on the complex coordinates $z = x+iy$ and $z^*$, which will allow us to make full use of the homogeneity and isotropy of $h$ later on.
In this new basis, we have
\begin{align}
  \frac{\partial}{\partial z}	& = \tfrac12 \frac{\partial}{\partial x} - \tfrac12i \frac{\partial}{\partial y}	& \frac{\partial}{\partial z^*}	& = \tfrac12 \frac{\partial}{\partial x} + \tfrac12i \frac{\partial}{\partial y}.
\end{align}
In this coordinate system, the definition of a maximum (minimum) becomes $h_z(\vec{r_0}) = 0$, $|h_{zz^*}(\vec{r_0})| > |h_{zz}(\vec{r_0})|$ and $h_{zz^*}(\vec{r_0})$ is negative (positive). \footnote{Note that $h_{zz^*}(\vec{r_0})$ is real valued.}

Some care is required however, since we are now dealing with \emph{complex} variables $h_z$ and $h_{zz}$ ($h_{zz^*}$ is real). We will treat the variables $z$ and $z^*$ as if they were independent. Therefore, next to $h_z$, we will consider $h_{z^*}$ as well, as a separate random variable, although it is actually the complex conjugate of $h_z$. Similarly, we also include $h_{z^*z^*} = h_{zz}^*$. Therefore, we are still dealing with five variables: $h_z$, $h_{zz}$, their conjugates, and $h_{zz^*}$.


As stated before, we will arrive at the joint probability distribution of these variables by building the generating function, which is the Fourier transform of the probability distribution. For a set of $n$ correlated variables $\xi_i$ this is
\begin{equation}
  \begin{split}
    & \chi(\lambda_1, \ldots, \lambda_n) \\
    & \qquad = \int \! \rd \xi_1 \ldots \rd \xi_n \, p(\xi_1, \ldots, \xi_n) e^{i(\xi_1\lambda_1 + \ldots + \xi_n\lambda_n)}.
  \end{split}
  \label{eq_gen_func}
\end{equation}
By expanding the exponential into a Taylor series we find that the coefficients -- which are called the \emph{moments} of the distribution (not to be confused with the moments from eq.~\eqref{eq_moments}) -- are correlations:
\begin{IEEEeqnarray}{rLl}
  & \mc{2}{l}{ \chi(\lambda_1, \ldots, \lambda_n) } \nl
  & \qquad =	& 1 + i \sum_j \la \xi_j \ra \lambda_j + \frac{i^2}{2!} \sum_{j_1,j_2} \la \xi_{j_1}\xi_{j_2} \ra \lambda_{j_1}\lambda_{j_2} \nl
  &		& +\: \frac{i^3}{3!} \sum_{j_1,j_2,j_3} \la \xi_{j_1}\xi_{j_2}\xi_{j_3} \ra \lambda_{j_1}\lambda_{j_2}\lambda_{j_3} + \ldots
  \label{eq_chi_moments}
\end{IEEEeqnarray}
If we do the same for the logarithm of $\chi$, we obtain the \emph{cumulants}:
\begin{align}
  \log \chi =\:	& i \sum_j C_1(\xi_j)\lambda_j + \frac{i^2}{2!} \sum_{j_1,j_2} C_2(\xi_{j_1},\xi_{j_2})\lambda_{j_1}\lambda_{j_2} \nl
  		& + \frac{i^3}{3!} \sum_{j_1,j_2,j_3} C_3(\xi_{j_1},\xi_{j_2},\xi_{j_3})\lambda_{j_1}\lambda_{j_2}\lambda_{j_3} + \ldots
  \label{eq_log_chi}
\end{align}
From eqs.~\eqref{eq_chi_moments} and ~\eqref{eq_log_chi} it can be derived that the cumulants can be factorized into moments, for example
\begin{equation}
  \begin{split}
    C_3(\xi_1, \xi_2, \xi_3) =\:	& \la \xi_1\xi_2\xi_3 \ra - \la \xi_1 \ra \la \xi_2\xi_3 \ra - \la \xi_2 \ra \la \xi_3\xi_1 \ra \\
    					& - \la \xi_3 \ra \la \xi_1\xi_2 \ra + 2 \la \xi_1 \ra \la \xi_2 \ra \la \xi_3 \ra.
  \end{split}
  \label{eq_cum_example}
\end{equation}
If all the cumulants are known, one can reconstruct the generating function and from that obtain the probability distribution via an inverse Fourier transformation.

The defining characteristic of Gaussian variables is that all cumulants are zero, apart from the second order ones ($C_2$). If $h$ were a Gaussian field, then this would apply to $p(h_z, h_{zz}, h_{zz^*})$, since the derivatives of a Gaussian field are themselves also Gaussian fields. Since $h$ is non-Gaussian, this is not the case. The first-order cumulants are still zero; for instance, we have $C(h_z) = \la h_z \ra = \partial_z \la h \ra = 0$ since $\la h \ra$ is constant due to the homogeneity of $h$. The third-order cumulants are however nonzero. We will include these and see how they influence the probability distribution and the densities of maxima and minima.

In principle, there are infinitely many nonzero cumulants. However, a field that is generated by a nonlinear differential equation, like eq.~\eqref{eq_diff_eqn}, typically has small cumulants of high order. In particular, if $f_{NL}$ is a quadratic function and the initial conditions are Gaussian, then the $n$-th order cumulants scale like $f_{NL}^{n-2}$ (for $n>2$) -- see appendix~\ref{app_cumulants}. Therefore we will only need to determine cumulants up to third order to get the correction to leading order.

The usefulness of the complex variables $z$ and $z^*$ becomes apparent when we look for all nonzero cumulants of second and third order involving the five variables we have. Since $h$ is isotropic, a moment like $\la h_{z^*} h_{zz} \ra$ should not change when we rotate the field by an arbitrary angle $\alpha$. Such a rotation would give $z \rightarrow e^{i \alpha}z$ and $z^* \rightarrow e^{-i \alpha}z^*$. Incorporating these in the derivatives causes the aforementioned moment to pick up a factor $e^{i \alpha}$. Since we argued that the moment should not be affected by the rotation, it must be zero. In general, any moment involving a different number of $z$ and $z^*$ derivatives is zero by this argument. Since cumulants can be decomposed into moments, as depicted in eq.~\eqref{eq_cum_example}, the same applies to cumulants.

Furthermore, translational symmetry implies some relations between the cumulants. From translational invariance it follows that any correlation should be constant with respect to $\vec{r}$. For instance, using the product rule, we have

\begin{equation}
  0  =  \partial_{z^*} \la h_z^2 h_{z^*} \ra  =  \la h_z^2 h_{z^*z^*} \ra + 2 \la h_z h_{z^*} h_{zz^*} \ra,
\end{equation}
which gives us the relation present in eq.~\eqref{eq_beta_def}.

Therefore, there are only a few independent cumulants that are (potentially) nonzero:
\begin{subequations}
  \begin{align}
    \sigma	& = \la |h_z|^2 \ra,  \label{eq_sigma_def} \\
    \alpha	& = \la |h_{zz}|^2 \ra = \la h_{zz^*}^2 \ra,  \label{eq_alpha_def} \\
    \beta	& = \la |h_z^2| h_{zz^*} \ra = -\tfrac12 \la h_z^2 h_{z^*z^*} \ra = -\tfrac12 \la h_{z^*}^2 h_{zz} \ra,  \label{eq_beta_def} \\
    \gamma	& = \la h_{zz^*}^3 \ra,  \label{eq_gamma_def} \\
    \delta	& = \la |h_{zz}|^2 h_{zz^*} \ra \label{eq_delta_def}.
  \end{align}
\end{subequations}
In these definitions, the cumulants have been expanded into moments in accordance with eq.~\eqref{eq_cum_example}; since the first-order correlations are zero, as noted before, only the third-order correlations remain. We also introduced the shorthand notation $|h_z|^2 = h_z h_z^* = h_z h_{z^*}$ and similarly for $|h_{zz}|^2$. Note also that the third-order cumulants, $\beta$, $\gamma$ and $\delta$ are close to zero when $h$ is close to being Gaussian, which we assume. On the other hand, $\sigma$ and $\alpha$ are nonzero in general.

We can now construct the logarithm of the generating function as prescribed by eq.~\eqref{eq_log_chi},
\begin{equation}
  \begin{split}
    \log \chi =	& -\sigma |\lambda_z|^2 - \alpha |\lambda_{zz}|^2 - \tfrac12 \alpha \lambda_{zz^*}^2 \\
    		& - i \beta |\lambda_z|^2 \lambda_{zz^*} + i \beta (\lambda_z^2 \lambda_{z^*z^*} + \lambda_{z^*}^2 \lambda_{zz}) \\
    		& - \tfrac{i}6 \gamma \lambda_{zz^*}^3 - i \delta |\lambda_{zz}|^2 \lambda_{zz^*}.
  \end{split}
\end{equation}
Note that some cumulants appear multiple times in eq.~\eqref{eq_log_chi} since the $\lambda$'s can be permuted (if they are not all the same); this explains why for instance the term $\lambda_{zz^*}^3$ has a prefactor $i/6$ whereas the prefactor of $|\lambda_{zz}|^2 \lambda_{zz^*} = \lambda_{zz}\lambda_{z^*z^*}\lambda_{zz^*}$ is $i$ (due to the 6 distinct permutations of the $\lambda$'s).

We see that $\chi$ features an exponential of a third-degree polynomial, making the inverse Fourier transform -- to be performed in order to get the probability distribution -- nontrivial. Remember however that the cubic terms are small owing to the near-Gaussianity of $h$, allowing us to make the expansion
\begin{equation}
  \begin{split}
    \chi =\:	& \Big[ 1 - i \beta |\lambda_z|^2 \lambda_{zz^*} + i \beta (\lambda_z^2 \lambda_{z^*z^*} + \lambda_{z^*}^2 \lambda_{zz}) \\
    		& \quad - \tfrac{i}6 \gamma \lambda_{zz^*}^3 - i \delta |\lambda_{zz}|^2 \lambda_{zz^*} \Big] \\
    		& \times \exp \big( \!-\! \sigma |\lambda_z|^2 - \alpha |\lambda_{zz}|^2 - \tfrac12 \alpha \lambda_{zz^*}^2 \big).
  \end{split}
\end{equation}
The inverse Fourier transform of this gives \footnote{A factor of $\pi^2$ rather than $(2\pi)^2$ is associated with the complex variables $h_z$ and $h_{zz}$ in the Fourier transform due to our normalization; see \cite{cite_paper2}.}
\begin{IEEEeqnarray}{rll}
  \mc{3}{l}{ p(h_z, h_{zz}, h_{zz^*}) } \nl
  \quad =\:	& \Big[	& 1 + \frac{\beta}{\alpha\sigma^2}h_{zz^*}(|h_z|^2 - \sigma) - \frac{\beta}{\alpha\sigma^2}(h_z^2h_{z^*z^*} + h_{z^*}^2h_{zz}) \nl
  		& 	& + \frac{\gamma}{6\alpha^3}(h_{zz^*}^3 - 3\alpha h_{zz^*}) + \frac{\delta}{\alpha^3}h_{zz^*}(|h_{zz}|^2-\alpha) \Big] \nl
  		& \mc{2}{l}{ \times \frac1{\pi^2\sqrt{2\pi}\sigma\alpha^{3/2}} e^{-|h_z|^2/\sigma - |h_{zz}|^2/\alpha - h_{zz^*}^2/2\alpha}. }
  \label{eq_jpd}
\end{IEEEeqnarray}

Now that the joint probability distribution of the relevant derivatives is obtained, we can set $h_z = h_{z^*} = 0$ -- this condition defines a critical point. The joint probability distribution measures how likely it is that $h_z$ and $h_{z^*}$ are \emph{close} to zero for a \emph{certain} point $\vec{r}$. What is needed however is for $h_z$ and $h_{z^*}$ to be \emph{exactly} zero for a point \emph{close} to $\vec{r}$, since we are looking for a density with respect to the $(x,y)$-plane. For this, we need to go from a probability density with respect to $h_z$ and $h_{z^*}$ to one with respect to $z$ and $z^*$ (representing $x$ and $y$). This is accomplished by multiplying $p$ with the following Jacobian:
\begin{equation}
  J  =  \bigg| \frac{\partial(h_z, h_{z^*})}{\partial(z, z^*)} \bigg|  =  \big| |h_{zz}|^2 - h_{zz^*}^2 \big|.
  \label{eq_Jacobian}
\end{equation}

Now we are ready to set $h_z = h_{z^*} = 0$ and integrate $pJ$ over $h_{zz}$ and $h_{zz^*}$. The range is determined by the type of critical point of interest; focus on the minima first. For these we must have $|h_{zz}| < |h_{zz^*}|$ and $h_{zz^*} > 0$. The integration over $h_{zz}$ is done by integrating over its real and imaginary part. Since the integrand depends only on the modulus of $h_{zz}$, we move to polar coordinates. Let us define $r = |h_{zz}|$ and $s = h_{zz^*}$. The integration range is then $0 < r < s$, and with eq.~\eqref{eq_jpd} we get
\begin{IEEEeqnarray}{rll}
  & \mc{2}{l}{ n_{min} =	\frac1{\pi^2\sqrt{2\pi}\sigma\alpha^{3/2}} } \nl
  & \quad \times \int_0^\infty \!	& \rd s \int_0^s \! 2\pi r \, \rd r \, (s^2 - r^2) e^{-r^2/\alpha - s^2/2\alpha} \label{eq_minima_int} \\
  &					& \times \Big[1 - \frac{\beta}{\alpha\sigma} s + \frac{\gamma}{6\alpha^3} (s^3-3\alpha s) + \frac{\delta}{\alpha^3} s(r^2-\alpha) \Big]. \nonumber
\end{IEEEeqnarray}
This integration is pretty straightforward: although the range of $r$ is finite, the integrand is a Gaussian multiplied by a polynomial that has only odd degrees of $r$, hence it does not give rise to error functions. The resulting integral over $s$ is also standard. The final result reads
\begin{equation}
  n_{min}  =  \frac{\alpha}{2\sqrt3\pi\sigma} - \frac1{\pi\sigma} \sqrt{\frac{\alpha}{2\pi}} \Big( \frac43 \frac{\beta}{\sigma} + \frac49 \frac{\delta}{\alpha} - \frac{10}{27} \frac{\gamma}{\alpha} \Big).
\end{equation}
For a Gaussian field, we would have $\beta = \gamma = \delta = 0$, $\sigma = \frac14 K_2$ and $\alpha = \frac1{16} K_4$. This would give us $n_{min} = K_4 / (8\sqrt3\pi K_2)$, exactly as given in \cite{cite_Longuet2}.

To get the density of maxima, the same integrand as in eq.~\eqref{eq_minima_int} needs to be integrated over the range $s < 0$ and $0 < r < -s$. However, note that if we make the transformation $s \rightarrow -s$, the range of integration is the same as in eq.~\eqref{eq_minima_int}. Furthermore, note that the transformation $s \rightarrow -s$ in the integrand is equivalent to $\beta \rightarrow -\beta$, $\gamma \rightarrow -\gamma$ and $\delta \rightarrow -\delta$. With this insight, we easily find that the expression for $n_{max}$ is the same as the above, except with a plus in place of the first minus.

With this result, the imbalance between maxima and minima is found to be
\begin{equation}
  \Delta n  \equiv  \frac{ n_{max} - n_{min} }{ n_{max} + n_{min} }  =  \sqrt{ \frac{6}{\pi \alpha} } \bigg( \frac43 \frac{\beta}{\sigma} + \frac49 \frac{\delta}{\alpha} - \frac{10}{27} \frac{\gamma}{\alpha} \bigg).
  \label{eq_maxmin}
\end{equation}

This is the main result of this paper. As an illustration, we shall now use this result to understand the evolution of maxima and minima in the context of a differential equation describing surface growth.

		\section{KPZ equation}
	\label{sec_kpz}

\begin{figure}
  \centering
  \includegraphics{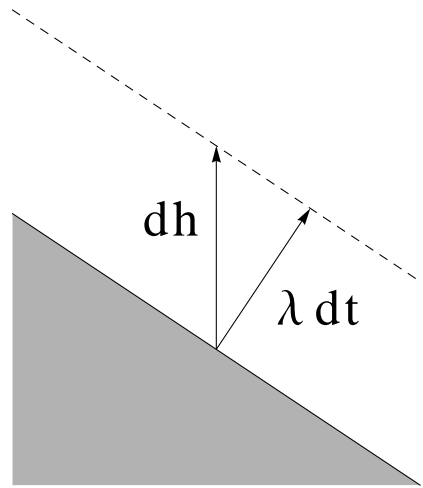}
  \caption{A geometrical interpretation of the KPZ equation applied to a growing surface. The surface is assumed to grow perpendicularly at a constant rate $\lambda$. Measured vertically, the growth rate is $\rd \kpzh / \rd t \approx \lambda(1 + \frac12 (\nabla \kpzh)^2)$.}
   \label{fig_kpz_figure}
\end{figure}

The deterministic Kardar-Parisi-Zhang (KPZ) equation \cite{cite_KPZ} is given by
\begin{equation}
  \frac{ \partial \kpzh }{ \partial t }  =  \nu \nabla^2 \kpzh + \frac{\lambda}{2} (\nabla \kpzh)^2.
  \label{eq_KPZ}
\end{equation}
This equation is often used to describe the height profile of a growing surface: the first term on the right-hand side describes the diffusion of particles along the surface, while the second term accounts for the assumption that the growth is perpendicular to the slope of the surface, while $\kpzh$ describes the height along the universal up direction \cite{cite_Barabasi}. This leads to (see fig.~\ref{fig_kpz_figure})
\begin{equation}
  \frac{\rd \kpzh}{\rd t}  =  \lambda \sqrt{1 + (\nabla \kpzh)^2}  =  \lambda + \frac{\lambda}{2} (\nabla \kpzh)^2 + \ldots
\end{equation}
The leading term $\lambda$ is ignored since it is just a constant that does not affect the profile of the surface.

Another interpretation of eq.~\eqref{eq_KPZ} is obtained by taking the gradient on both sides, which yields
\begin{equation}
  \frac{ \partial \vec{v} }{ \partial t }  =  \nu \nabla^2 \vec{v} + \lambda \vec{v} \vec{\nabla} \vec{v},
\end{equation}
where $\vec{v} = \vec{\nabla} \kpzh$ is a velocity field. This is a vector Burger's equation which arises in fluid mechanics. The maxima and minima of $\kpzh$ correspond to sources and sinks of $v$.

We will take $\kpzh(\vec{r}, t)$ to be a Gaussian field at $t=0$, and use our result eq.~\eqref{eq_maxmin} to determine how the non-Gaussianities, which arise and evolve due to the KPZ equation, influence the densities of maxima and minima.

First note that if we would set $\lambda = 0$ in eq.~\eqref{eq_KPZ}, we retrieve the heat equation, which preserves the Gaussianity of a field: if we enter $\kpzh(\vec{r}, t=0) = H(\vec{r})$, where $H(\vec{r})$ is a Gaussian field as given by eq.~\eqref{eq_gaussian}, we find that the solution is
\begin{equation}
  \kpzh(\vec{r}, t)  =  \sum_{\vec{k}} A(k) e^{-k^2 \nu t} \cos(\vec{k} \cdot \vec{r} + \phi_k).
  \label{eq_heatsol}
\end{equation}
We find that the amplitudes pick up a factor $\exp(-k^2 \nu t)$, but the phases remain independent. Therefore, even though its amplitude spectrum changes, $\kpzh(\vec{r},t)$ remains Gaussian at any time $t$ and the density of maxima and minima remains the same, since this is a general property of Gaussian fields.

If we have $\lambda \neq 0$, $\kpzh(\vec{r}, t)$ no longer remains Gaussian. In fact, as we will see, the density of maxima and minima is no longer the same. We shall assume $\lambda$ to be small in comparison with $\nu$, and find out how these densities differ as a function of time, using eq.~\eqref{eq_maxmin}. For this, we need to determine the two- and three-point correlations $\sigma$, $\alpha$, $\beta$, $\gamma$ and $\delta$.

First, we substitute $u = \exp((\lambda/2\nu) \kpzh)$. Note that, since this is a monotonically increasing function of $\kpzh$, the maxima and minima of $u$ are exactly the same points as those of $\kpzh$. In terms of $u$, the KPZ equation becomes:
\begin{equation}
  \frac{ \partial u}{ \partial t }  =  \nu \nabla^2 u,
  \label{eq_kpz_heat}
\end{equation}
which is simply the heat equation. However, $u(\vec{r}, t=0) = \exp((\lambda/2\nu) \kpzh_0)$ is now not a Gaussian field. If we assume that $\lambda \ll \nu$, we have:
\begin{equation}
  u_0  =  1 + \frac{\lambda}{2\nu} \kpzh_0 + \frac{\lambda^2}{8\nu^2} \kpzh_0^2 + O((\lambda/\nu)^3).
\end{equation}
Since the leading term, equal to one, has no influence on either the maxima and minima or the heat equation, we can ignore it. The same applies to the prefactor $\frac{\lambda}{2\nu}$ of the second term. Hence we make a final transformation
\begin{equation}
  v  \equiv  \frac{2\nu}{\lambda}(u-1),
\end{equation}
\begin{equation}
  v_0  =  \kpzh_0 + \frac{\lambda}{4\nu} \kpzh_0^2 + O((\lambda/\nu)^2).
\end{equation}
Note that $v$ still obeys the heat equation and also shares the same maxima and minima with $\kpzh$ and $u$. Moreover, we now have $v(\vec{r}, t=0)$ in the desired form of a Gaussian $\kpzh_0$ plus a perturbation. Since $v$ obeys the heat equation, we can use the corresponding Green's function to write down the general solution
\begin{align}
  v(r, t)	& =  \int \rd^2 \tilde{r} \, G(r, \tilde{r}, t) v_0(\tilde{r}) \nl
  		& =  \int \rd^2 \tilde{r} \, \frac1{4 \pi \nu t} e^{ -\frac{ (r-\tilde{r})^2 }{ 4 \nu t } } \big( \kpzh_0(\tilde{r}) + \frac{\lambda}{4\nu} \kpzh_0(\tilde{r})^2 \big),
  \label{eq_greensfunc}
\end{align}
where $v_0(\tilde{r}) = v(r, t=0)$.

We can now calculate the five correlations needed to determine $\Delta n$. We will demonstrate the procedure using $\sigma = \la v_z(r,t) v_{z^*}(r,t) \ra$ as an example.
\begin{IEEEeqnarray}{rLl}
  \sigma	& \mc{2}{L}{ = \la v_z(r,t) v_{z^*}(r,t) \ra } \nl
  		& = \iint & \rd^2 \tilde{r}_1 \rd^2 \tilde{r}_2 \, \partial_{z_1} G(r_1,\tilde{r}_1,t) \nl
  		&	  & \partial_{z^*_2} G(r_2,\tilde{r}_2,t) \la v_0(\tilde{r}_1) v_0(\tilde{r}_2) \ra \Big|_{r_1=r_2=r}.
  \label{eq_sigma}
\end{IEEEeqnarray}
The brackets represent averaging over all $\phi_{\vec{k}}$ that define $v_0$, while the spatial derivatives act only on the respective Green's function. The latter gives
\begin{align}
  & \partial_{z_1} G(r_1,\tilde{r}_1,t)	 =  \partial_{z_1} \Big( \frac1{4 \pi \nu t} e^{ -\frac{ (r_1-\tilde{r}_1)^2 }{ 4 \nu t } } \Big) \nl
  & \qquad =  \frac1{\pi(4 \nu t)^2} ((x_1-\tilde{x}_1) - i(y-\tilde{y}_1)) e^{ -\frac{ (r_1-\tilde{r}_1)^2 }{ 4 \nu t } }.
  \label{eq_green_deriv}
\end{align}
The moment present in eq.~\eqref{eq_sigma} is
\begin{IEEEeqnarray}{rLl}
  & \mc{2}{l}{ \la v_0(\tilde{r}_1) v_0(\tilde{r}_2) \ra } \nl
  & \qquad =	& \big\la \big( \kpzh_0(\tilde{r}_1) + \frac{\lambda}{4\nu} \kpzh_0(\tilde{r}_1)^2 \big) \big( \kpzh_0(\tilde{r}_2) + \frac{\lambda}{4\nu} \kpzh_0(\tilde{r}_2)^2 \big) \big\ra \nl
  & \qquad =	& \la \kpzh_0(\tilde{r}_1) \kpzh_0(\tilde{r}_2) \ra \nl
  &		& +\: \frac{\lambda}{4\nu} \big( \la \kpzh_0(\tilde{r}_1) \kpzh_0(\tilde{r}_2)^2 \ra + \la \kpzh_0(\tilde{r}_1)^2 \kpzh_0(\tilde{r}_2) \ra \big) \nl
  &		& +\: \Big( \frac{\lambda}{4\nu} \Big)^2 \la \kpzh_0(\tilde{r}_1)^2 \kpzh_0(\tilde{r}_2)^2 \ra.
\end{IEEEeqnarray}
Note that the second term (the one linear in $\lambda/4\nu$) is a three-point correlation, and therefore zero due to the symmetry of the Gaussian field $\kpzh_0$. We will ignore the last term since our analysis is restricted to first order in $\lambda/4\nu$. All that remains is the two-point correlation, which with the help of eq.~\eqref{eq_gaussian} is seen to be
\begin{align}
  \la v_0(\tilde{r}_1) v_0(\tilde{r}_2) \ra	& = \la \kpzh_0(\tilde{r}_1) \kpzh_0(\tilde{r}_2) \ra \nl
  						& = \sum_{\vec{k}} \tfrac12 A(k)^2 \cos(\vec{k} \cdot (\tilde{r}_1 - \tilde{r}_2)).
  \label{eq_corr}
\end{align}
We will now plug our intermediate results, eqs.~\eqref{eq_green_deriv} and \eqref{eq_corr}, back into eq.~\eqref{eq_sigma}. For convenience, we will set $\vec{r} = \vec{0}$, which we are allowed to do thanks to the homogeneity of $v$. We find
\begin{equation}
  \begin{split}
    \sigma  =  \sum_{\vec{k}} \tfrac12 A(k)^2 \iint	& \rd^2\tilde{r}_1 \rd^2\tilde{r}_2 \, \pi^{-2}(4\nu t)^{-4} (\tilde{r}_1 \cdot \tilde{r}_2) \\
    							& e^{-(\tilde{r}_1^2 + \tilde{r}_2^2)/(4\nu t)} \cos(\vec{k} \cdot (\tilde{r}_1 - \tilde{r}_2)).
  \end{split}
  \label{eq_sigma_int}
\end{equation}
Note that based on eq.~\eqref{eq_green_deriv} we should have put $(\tilde{x}_1-i\tilde{y}_1)(\tilde{x}_2+i\tilde{y}_2)$ instead of $(\tilde{r}_1 \cdot \tilde{r}_2)$; the latter is merely the real part of the former. However, since we already know that the final answer is real (since $\sigma = \la |v_z|^2 \ra$), we can conclude that the imaginary part would not give a contribution.

After performing the integrals in eq.~\eqref{eq_sigma_int} we get the result given below. The three-point correlations $\beta$, $\gamma$ and $\delta$ give rise to six-dimensional integrals involving four-point correlations (which are first order in $\lambda/4\nu$). These correlations can be factorized into two two-point correlations by Wick's theorem, resulting in a sum over two wave vectors $\vec{k}_1$ and $\vec{k}_2$, as opposed to the one we had in the case of $\sigma$.

All the relevant correlations are
\begin{IEEEeqnarray}{rLl}
  \sigma	& \mc{2}{L}{ = \sum_{\vec{k}} \frac12 A(k)^2 \frac14 k^2 e^{-2 k^2 \nu t}, }  \IEEEyessubnumber \\
  \alpha	& \mc{2}{L}{ = \sum_{\vec{k}} \frac12 A(k)^2 \frac1{16} k^4 e^{-2 k^2 \nu t}, }  \IEEEyessubnumber \\
  \beta		& = -\frac{\lambda}{4\nu} \sum_{\vec{k}_1} \sum_{\vec{k}_2}	& \frac14 A(k_1)^2 A(k_2)^2 \frac14 [k_1^2 k_2^2 - (\vec{k}_1 \cdot \vec{k}_2)^2]  \nl
  		&								& \times e^{ -2(k_1^2 + k_2^2 + \vec{k_1} \cdot \vec{k_2}) \nu t },  \IEEEyessubnumber \\
  \gamma	& = -\frac{\lambda}{4\nu} \sum_{\vec{k}_1} \sum_{\vec{k}_2}	& \frac14 A(k_1)^2 A(k_2)^2 \frac3{32} k_1^2 k_2^2 (\vec{k}_1+\vec{k}_2)^2  \nl
  		&								& \times e^{ -2(k_1^2 + k_2^2 + \vec{k_1} \cdot \vec{k_2}) \nu t },  \IEEEyessubnumber \\
  \delta	& = -\frac{\lambda}{4\nu} \sum_{\vec{k}_1} \sum_{\vec{k}_2}	& \frac14 A(k_1)^2 A(k_2)^2 \frac1{32} \Big[ \!-\! k_1^2 k_2^2 (k_1^2+k_2^2)  \nl
  		&								& +\: ((\vec{k}_1+\vec{k}_2)^4-k_1^4-k_2^4)(\vec{k}_1 \cdot \vec{k}_2) \Big]  \nl
  		&								& \times e^{ -2(k_1^2 + k_2^2 + \vec{k_1} \cdot \vec{k_2}) \nu t }.  \IEEEyessubnumber
\end{IEEEeqnarray}
For a continuous spectrum, the sums can be replaced by integrals.

\begin{figure}
  \centering
  \subfigure[]{\includegraphics[width=0.9\columnwidth]{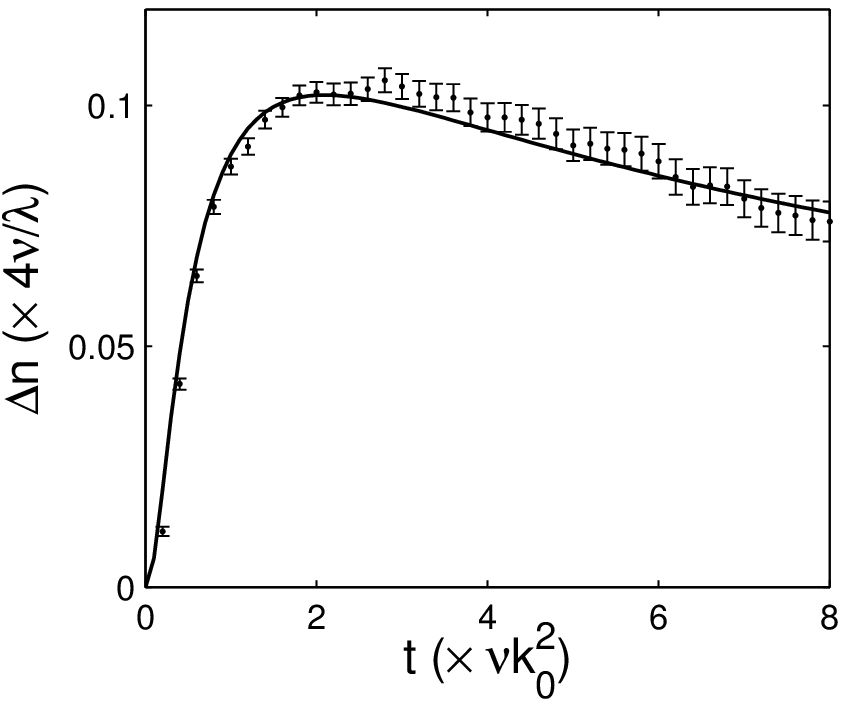} \label{fig1}}
  \subfigure[]{\includegraphics[width=0.9\columnwidth]{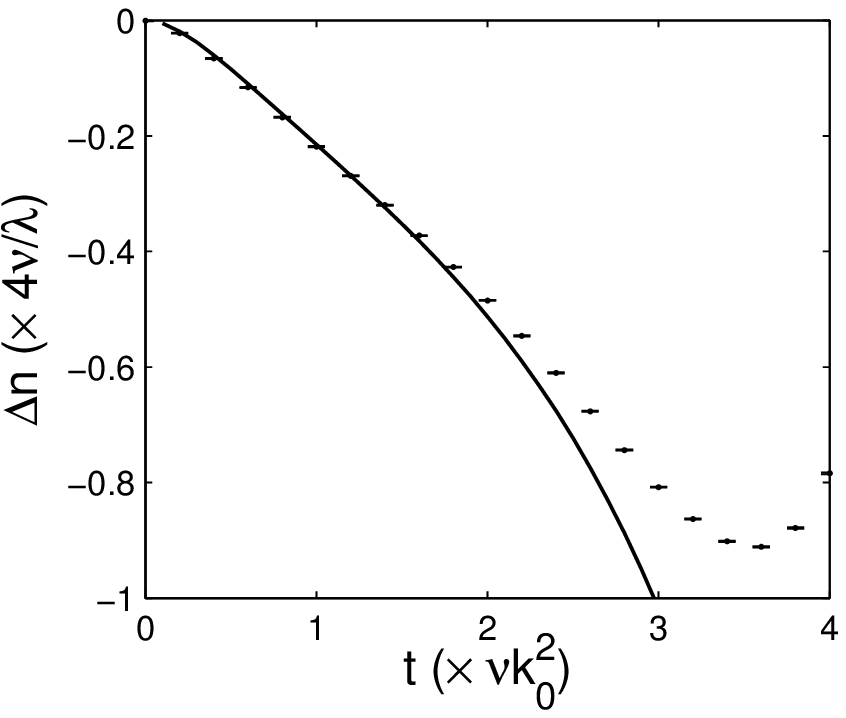} \label{fig2}}
  \caption{The imbalance between maxima and minima $\Delta n$ of $\kpzh(\vec{r}, t)$, where $\kpzh$ obeys the KPZ equation (with $\lambda/4\nu = 0.1$), as a function of time. At $t=0$, $\kpzh(\vec{r})$ was taken to be a Gaussian field with (a) a Gaussian spectrum $A(k) \propto \exp(-k^2/(4k_0^2))$; (b) a ring spectrum $A(k) \propto \delta(k-k_0)$. Shown are our theoretical perturbative result (eq.~\eqref{eq_maxmin}) and data from simulations.}
\end{figure}

We see that the parameters depend on the spectrum of $\kpzh_0$ in a nontrivial way. Especially the presence of $\vec{k_1} \cdot \vec{k_2}$ (which is also present in terms like $(\vec{k_1}+\vec{k_2})^2$) in the relations for $\beta$, $\gamma$ and $\delta$ complicates matters, as it introduces a dependence on the angle between $\vec{k_1}$ and $\vec{k_2}$. An exact analytical evaluation is therefore only realizable for a few spectra of a convenient form. Even for the so-called ring spectrum, with $A(k) \propto \delta(k-k_0)$, arguably the simplest spectrum one can have, the angular dependence introduces nontrivial functions. In this case, eq.~\eqref{eq_maxmin} reads
\begin{equation}
  \begin{split}
    \Delta n  =  \frac{\lambda}{4\nu} \frac89 \sqrt{\frac6{\pi}} \frac{e^{-\tau}}{\tau} \Big(	& \!-\! (2+\tau) I_0(2\tau) - 5\tau I_1(2\tau) \\
    												& + (2+\tau+6\tau^2) {}_{0\!}F_1(2; \tau^2) \Big),
  \end{split}
  \label{eq_maxmin_ringspec}
\end{equation}
where $\tau \equiv k_0^2 \nu t$; $I_0$ and $I_1$ are modified Bessel functions of the first kind and ${}_{0\!}F_1$ is the confluent hypergeometric function.
Recall that we set $K_0 = \la \kpzh_0^2 \ra = 1$ for convenience; for the general case, a factor of $\sqrt{K_0}$ needs to be added.

Another, more elegant case in which an exact evaluation of eq.~\eqref{eq_maxmin} is possible is the Gaussian spectrum $A(k) \propto \exp(-k^2/(4k_0^2))$, for which
\begin{equation}
  \Delta n  =  \frac{\lambda}{4\nu} \frac{ 64 \tau^3 (1+4\tau)^{7/2} }{ \sqrt{3\pi} (1+2\tau)^3 (1+6\tau)^4 },
  \label{eq_maxmin_gsnspec}
\end{equation}
where again $\tau \equiv k_0^2 \nu t$ and a factor of $\sqrt{K_0}$ needs to be added for our result to apply in general.

Going back to the general case of an unspecified power spectrum, it is convenient to expand $\Delta n$ in $t$. The result is
\begin{equation}
  \Delta n  =  \frac{\lambda}{4\nu} \frac49 \sqrt{\frac6{\pi}} \frac{2 K_2 K_6 - 3 K_4^2}{K_2 \sqrt{K_4}} (\nu t)^2 + O(t^3),
  \label{eq_maxmin_expansion}
\end{equation}
for all $K_0$. One may note that for a Gaussian spectrum, there is no quadratic order in eq.~\eqref{eq_maxmin_gsnspec}, which is confirmed by the above formula, since $2 K_2 K_6 - 3 K_4^2 = 0$ in this case.

The analytical results for $\Delta n$ above are compared to results from numerical simulations (with $K_0 = 1$ and $\lambda/4\nu = 0.1$) in figs.~\ref{fig1} and \ref{fig2}. The general method is the same as outlined in ref.~\cite{cite_paper1}. We start with a Gaussian field $\kpzh_0$ defined on a finite square grid with periodic boundary conditions. We then transform to $v_0$ and use the alternating direction implicit (ADI) method to simulate the heat equation, collecting statistics on the maxima and minima at every time step. The results are averaged over for tens of thousands of $\kpzh_0$'s, each with the same spectrum but random phases.

In general, if a field evolves under a nonlinear equation for a long time, the non-Gaussianity can become large, even when the perturbation is small, because it will add up over time. Thus we may expect a breakdown of our predictions after some time, as in fig.~\ref{fig2}. However, the KPZ equation has a special mapping to a diffusion equation (eq.~\eqref{eq_kpz_heat}), and this implies that the non-Gaussian perturbations never build up. Eq.~\eqref{eq_greensfunc} shows that the nonlinear correction diffuses outward but does not grow over time. Therefore, for the KPZ equation, our approximations should remain accurate for arbitrarily long times. This is indeed what we see in fig.~\ref{fig1}, where $\kpzh(t=0)$ is Gaussian field with a Gaussian spectral function.

In fig.~\ref{fig2} however there is a breakdown for the ring spectrum. This spectrum is special because it has zero weight at $k=0$. This implies that the leading Gaussian term in eq.~\eqref{eq_greensfunc} is suppressed exponentially, decaying as $\exp(-k_0^2 \nu t)$ (see eq.~\eqref{eq_heatsol}). Thus after a long time, the second term dominates, and our approximation that $v$ is close to a Gaussian no longer holds. Whenever the spectral function has a weight at $k=0$ (as in fig.~\ref{fig1}), the approximation works for a longer time.

		\section{Conclusions}
	\label{sec_conclusions}

We have found a general perturbative formula, eq.~\eqref{eq_maxmin}, for determining the imbalance between maxima and minima of an isotropic random field that is almost Gaussian. It allows one to attack the reverse problem, namely, to determine the size of the phenomenon that causes the non-Gaussianity, by measuring the relative densities of maxima and minima. In the case of the deterministic KPZ equation for instance, the imbalance can reveal the size of the nonlinear parameter $\lambda$ relative to the diffusion coefficient $\nu$.

In ref.~\cite{cite_paper1}, we investigated the imbalance between maxima and minima as a result of non-Gaussianity. Although we arrived at an exact result, it applied only to the special case of a local perturbation, i.e.\ for a field given by $h(\vec{r}) = F_{NL}(H(\vec{r}))$ where $H$ is a Gaussian field and $F_{NL}$ any (nonlinear) function. The result in the present study, although perturbative, also accommodates \emph{nonlocal} perturbations, provided that the resulting field is still homogeneous and isotropic.

For local perturbations, we found that the size of the imbalance is exponentially small in the size of the perturbation \cite{cite_paper1}. Nonlocal perturbations however allow for a power-law relation. This is apparent in eq.~\eqref{eq_maxmin_expansion}, which shows that the KPZ equation can cause an imbalance that grows quadratically with time. As a result, the densities of maxima and minima can prove to be a sensitive test to not only detect non-Gaussianity, but also to distinguish local from nonlocal perturbations that induce non-Gaussian statistics.

\begin{acknowledgments}
This work was supported by the Dutch Foundation for Fundamental Research on Matter (FOM), the Dutch Foundation for Scientific Research (NWO) and the European Research Council (ERC).
We thank T. Lubensky, R.D. Kamien, B. Jain, A. Boyarsky, L. Mahadevan, B. Chen and W. van Saarloos for stimulating discussions.
\end{acknowledgments}

\appendix
		\section{Higher order cumulants}
	\label{app_cumulants}

In this section it is demonstrated that, for an initially Gaussian field evolving according to a diffusion equation with a perturbative nonlinear term, the cumulants become smaller as the order increases (i.e.\ they are of higher order in the perturbation).

Consider the equation
\begin{equation}
  \dot{h}_n  =  \sum_m A_{nm} + \eps \sum_{p,q} B_{npq} h_p h_q,
  \label{eq:silo}
\end{equation}
with the initial condition
\begin{equation}
  h_n(0)  =  H_n,
  \label{eq:corn}
\end{equation}
where the $H_n$'s are a set of variables with a joint Gaussian distribution. These coupled differential equations are a simple model of a nonlinearity, with the lowest order (quadratic), and they also include the KPZ equation as a special case, if it is discretized. This differential equation illustrates the general principle that cumulants of a high order are very small if the nonlinear term in the differential equation is small -- unless one waits long enough for these cumulants to build up.

For this family of equations the precise result is that, after a finite period of time, the $k$-th order cumulants of any of the $h_n$'s are of order at most $\eps^{k-2}$ if $k>2$ (for $k=1$ or $k=2$ they are bounded).

There are two steps in the proof: first, we find how $h_n$ depends on the initial conditions, and show that it has the form of a power series in $\eps$. The result is that
\begin{equation}
  \begin{split}
    h_n(t) =\:	& F_n^{(0)}(\{H_j\}) + \eps F_n^{(1)}(\{H_j\}) \\
    		& + \eps^2 F_n^{(2)}(\{H_j\}) + \ldots
  \end{split}
  \label{eq:train}
\end{equation}
where $F_n^{(0)}$ is a linear function, $F_n^{(1)}$ is quadratic, etc. So the dependence of a given term on the $H_j$'s is polynomial; the dependence on $t$ is all in the coefficients of these polynomials.

In other words, $h_n$ can be expressed in the form of a nonlinear function of a Gaussian, the same type of function whose cumulants we calculated in \cite{cite_paper2}. We will see that many of the cumulants vanish -- this is the second step of the proof. We calculate the cumulants,
\begin{equation}
  C_k(h_{n_1},\ldots, h_{n_k})  =  \sum_{r=0}^\infty \eps^r \!\! \sum_{\substack{r_1,r_2,\ldots,r_k \\ \sum r_i=r}} \!\!\! C_k(F_{n_1}^{(r_1)}, \ldots, F_{n_k}^{(r_k)}).
  \label{eq:solitaire}
\end{equation}
All the terms up to order $r=k-3$ vanish, so that the remaining terms are of order $\eps^{k-2}$ or smaller. This is a consequence of a general theorem: a cumulant of $k$ polynomials in Gaussian variables is zero if 
\begin{equation}
  k>1+\frac{d}{2}.
  \label{eq:mushroom}
\end{equation}
where $d$ is the sum of the degrees of the polynomials. In $C_k(F_{n_1}^{(r_1)},\ldots,F_{n_k}^{(r_k)})$ the sum of the degrees is $d=\sum_i r_i+1=r+k$. If $r\leq k-3$, then Eq. (\ref{eq:mushroom}) follows, so the cumulant vanishes.

\subsection{Power series solution}
Expand $h_n(t)=\sum_r \eps^r h_n^{(r)}(t)$ and substitute it into eq.~\eqref{eq:silo}, and then match the coefficients of $\eps^r$. This gives the relation
\begin{equation}
  \begin{split}
    & \frac{\partial}{\partial t} h_n^{(r)}(t) - \sum_m A_{nm} h_m^{(r)}(t) \\
    & \quad = \sum_{p,q} \sum_{r_1=0}^{r-1} B_{npq} h_p^{(r_1)}(t) h_q^{(r-1-r_1)}(t).
  \end{split}
\end{equation}
Here, everything depending on $h^{(r)}$ is on the left-hand side; everything on the right-hand side depends on earlier terms in the series, $h^{(r_1)}$ with $r_1<r$. This means that one can solve the equations recursively: first find the $h$'s up to $r_1=r-1$, then substitute it into the right-hand side of the equation and then solve for $h^{(r)}$, which is straightforward because it is a linear equation with a source. We only need to know the initial conditions, which are
\begin{equation}
  h^{(0)}_n=H_n; \quad h^{(r)}_n = 0 \text{ for } r \geq 1.
\end{equation}
The solutions to the equations are given as follows:
\begin{equation}
  h_n^{(0)}(t)  =  \sum_m \big[ e^{At} \big]_{nm} H_m,
\end{equation}
\begin{equation}
  \begin{split}
    h_n^{(r)}(t) = \int_0^t dt' \sum_{m,p,q} \sum_{r_1 = 0}^{r-1}	& \big[ e^{A(t-t')} \big]_{nm} B_{mpq} \\
    									& \times h_p^{(r_1)}(t') h_q^{(r-1-r_1)}(t'),
  \end{split}
  \label{eq:powerseriessol}
\end{equation}
where $e^{At}$ is the exponential of the matrix $At$, which is just a set of functions of $t$.

These functions are all polynomials in the $H_j$'s. First, $h_n^{(0)}$ is obviously linear. Entering $r=1$ in eq.~\eqref{eq:powerseriessol} shows that $h_n^{(1)}$ is the sum and integral of $h_p^{(0)} h_q^{(0)}$, which is thus quadratic in the $H_j$'s. Now we can find the general dependence inductively: assume that it has already been shown that $h_n^{(r_1)}$ is a degree $r_1+1$ polynomial in the $H_j$'s for $r_1<r$. Then $h_p^{(r_1)}(t') h_q^{(r-1-r_1)}(t')$ is of degree $r+1$, and thus $h^{(r)}$ is as well.

\subsection{Vanishing cumulants}

We will calculate the cumulants of polynomials in the $H_j$'s by reducing them to cumulants of the $H_j$'s themselves, which are Gaussian. A helpful identity for this expresses $C(xy,z_1,\ldots,z_q)$ where $x,y,z_i$ are any random variables in terms of simpler cumulants. The identity is
\begin{equation}
  \begin{split}
    C(xy, z_1, \ldots, z_q) =\:	& C(x, y, z_1, \ldots, z_q) \\
    				& + \hspace{-15pt} \sum_{S \cup T = \{1,\ldots,q\}} \hspace{-15pt} C(x,z_S) C(y,z_T).
  \end{split}
\label{eq:divisions}
\end{equation}
The sum is over all ways of partitioning the indices of the $z$'s into two sets $S$ and $T$. The symbol $z_S$ is short for the list of all the $z$'s corresponding to the indices $S$.

Here is an example:
\begin{equation}
  \begin{split}
    C(xy,u,v) =\:	& C(x,y,u,v) + C(x)C(y,u,v) \\
    			& + C(x,u)C(y,v) + C(x,v)C(y,u) \\
    			& + C(x,u,v)C(y).
  \end{split}
\end{equation}

A proof of this relation can be obtained using induction. First note that it is trivially true for $q=0$, since $C(x,y) = \la xy \ra - \la x \ra \la y \ra$. Now we assume the relation to hold for all $q' < q$. Consider the identity (see e.g.\ \cite{cite_paper2} or \cite{cite_Kampen})
\begin{equation}
  \la x_1 \ldots x_n \ra  =  \sum C(x_{S_1}) C(x_{S_2}) \ldots C(x_{S_m}),
\end{equation}
where the sum is taken over all the ways in which the set $\{1,\ldots,n\}$ can be partitioned into disjoint subsets $S_i$. If we apply this identity to the set $\{x,y,z_1,\ldots,z_q\}$ and group together the terms for which $x$ and $y$ are in the same subset or in different ones, we find
\begin{IEEEeqnarray}{lLll}
  & \mc{3}{l}{ \la x y z_1 \ldots z_n \ra } \nl
  & \quad =	& \mc{2}{l}{ \sum_{U,\{V_i\}} C(x,y,z_U) C(z_{V_1}) \ldots C(z_{V_m}) } \nl
  &		& \mc{2}{l}{ + \!\! \sum_{S,T,\{V_i\}} \!\! C(x,z_S) C(y,z_T) C(z_{V_1}) \ldots C(z_{V_m}) } \nl
  & \quad =	& \sum_{U,\{V_i\}} C(z_{V_1}) \ldots C(z_{V_m}) \Big[	& C(x,y,z_U) \\
  &		&							& + \!\! \sum_{S \cup T = U} \!\! C(x,z_S) C(y,z_T) \Big] \nonumber
\end{IEEEeqnarray}
We can also choose to expand $\la x y z_1 \ldots z_n \ra$ while treating $xy$ as a single variable, which results in
\begin{equation}
  \la x y z_1 \ldots z_n \ra  =  \sum_{U,\{V_i\}} C(xy,z_U) C(z_{V_1}) \ldots C(z_{V_m})
\end{equation}
The two decompositions into cumulants should be equal. By induction, we can pose
\begin{equation}
  C(xy,z_U)  =  C(x,y,z_U) + \!\! \sum_{S \cup T = U} \!\! C(x,z_S) C(y,z_T)
\end{equation}
for all $U \neq \{1,\ldots,q\}$. It then easily follows that the relation must also hold for $U = \{1,\ldots,q\}$.

We will use this identity to prove that if $p_1,\ldots p_k$ are degree $d_1,\ldots d_k$ polynomials in Gaussian variables and $d=\sum_i d_i$, then $C_k(p_1,\ldots,p_k)$ vanishes if eq.~\eqref{eq:mushroom} is satisfied. We shall first demonstrate the procedure using a simple example: $C(H^2,H^2,H,H,H)$ where $H$ is a Gaussian variable. We will reduce this to cumulants of $H$ by using eq.~\eqref{eq:divisions}; that will mean we have to apply the identity twice to split up both $H^2$'s. After the first time, we have a sum featuring one term with a single cumulant, $C(H^2,H,H,H,H,H)$, while the other terms are products of two cumulants. Furthermore, there is only one $H^2$ left in each term. After applying eq.~\eqref{eq:divisions} a second time, we are left with products of at most three cumulants. Since there are $7$ $H$'s distributed among these cumulants, at least one of the cumulants in each product is of at least third order, and hence zero because the $H$'s are Gaussian. Hence $C(H^2,H^2,H,H,H) = 0$.

In general, we first use the multilinear property of the cumulant function (i.e.\ $C(x+y,z,w,\ldots)=C(x,z,w,\ldots)+C(y,z,w,\ldots)$) to reduce each of the variables to one term (which is a product of some of the $H$'s). It takes $d-k$ applications of eq.~\eqref{eq:divisions} to split all the variables up into individual $H$'s, because it takes $d_i-1$ steps to factor the $i$-th variable, for a total of $\sum_i d_i-1=d-k$ steps.
Since each application of eq.~\eqref{eq:divisions} adds at most one cumulant to each term, in the end each term has at most $d-k+1$ factors of $C$. This is less than $\frac{d}{2}$ by eq.~\eqref{eq:mushroom}. But there are a total of $d$ variables $H$'s that are split among them. Hence one of the factors is a third-order cumulant or higher, which means that it has to be zero.

Now this result can be combined with eq.~\eqref{eq:train} to prove that the $k$-th order cumulants of the $h_n$'s are of order $\eps^{k-2}$, as we showed above.

\bibliography{nongauss}

\end{document}